\documentclass[twocolumn,prl]{revtex4}

\usepackage{dcolumn}
\usepackage{amsmath}

\usepackage{graphicx}
\usepackage{pslatex}

\newcommand{\kin}{k^\mathrm{in}}
\newcommand{\kout}{k^\mathrm{out}}

\newcommand{\kappain}{\kappa^\mathrm{in}}
\newcommand{\kappaout}{\kappa^\mathrm{out}}

\renewcommand{\d}{\mathrm{d}}
\newcommand{\e}{\mathrm{e}}

\newcommand{\Ord}{\mathrm{O}}

\newlength{\figurewidth}
\begin{document}

\title{Random acyclic networks}
\author{Brian Karrer and M. E. J. Newman}
\affiliation{Department of Physics, University of Michigan, Ann Arbor, MI
  48109}
\affiliation{Santa Fe Institute, 1399 Hyde Park Road, Santa Fe, NM 87501}
\begin{abstract}
  Directed acyclic graphs are a fundamental class of networks that includes
  citation networks, food webs, and family trees, among others.  Here we
  define a random graph model for directed acyclic graphs and give
  solutions for a number of the model's properties, including connection
  probabilities and component sizes, as well as a fast algorithm for
  simulating the model on a computer.  We compare the predictions of the
  model to a real-world network of citations between physics papers and
  find surprisingly good agreement, suggesting that the structure of the
  real network may be quite well described by the random graph.
\end{abstract}
\maketitle

Many networks of scientific interest take the form of directed acyclic
graphs---directed networks containing no closed cycles, i.e.,~paths that
start and end at the same vertex and follow edges only in the forward
direction~\cite{Newman03d}.  The best known examples are citation
networks~\cite{Price65} but there are many others as well, such as family
trees, phylogenetic networks, food webs, feed-forward neural networks, and
software call graphs.  (Some of these are only approximately acyclic, but
the approximation is typically good enough that acyclic graphs still
provide a useful starting point for theories of network structure.)

One of the most fundamental and important of theoretical models in the
study of networks is the random graph.  In its most general form, a random
graph is a model network of a given number of vertices in which certain
topological features are fixed but in all other respects edges are placed
at random~\cite{ER59,Bollobas01,Bollobas80,MR95,NSW01}.  Random graphs have
significant advantages as models of networks, allowing one to isolate the
effects of particular structural parameters and being exactly solvable for
many of their topological properties, both local and global.  They have
played a central role in the development of network theory, proving useful
as a guide to both the qualitative and the quantitative properties of
networks of many kinds.

In this Letter, we present a random graph model for directed acyclic
graphs.  Despite the name ``acyclic graph,'' the lack of cycles is in fact
not the defining feature of most real-world acyclic graphs.  The defining
feature is that the vertices have a natural ordering.  In a citation
network of scientific papers, for instance, the papers are time-ordered by
publication date and the network is acyclic because papers can only cite
those that came before them, meaning that all edges point backward in time.
(Note that self-edges are not allowed in acyclic graphs.)  It is clear that
all networks ordered in this way are acyclic, and it can be proved that for
all acyclic networks at least one appropriate ordering of the vertices
exists.  In practical situations, however, the ordering is normally the
crucial property and it will be the defining feature for the models
described in this paper.

Suppose then that we are given an ordered set of $n$ vertices denoted by
$i=1\ldots n$ and a corresponding degree sequence, i.e.,~a complete set of
in- and out-degrees $\kin_i$ and~$\kout_i$ for all vertices.  In our
representation all edges will point from ``later'' vertices (higher~$i$) to
``earlier'' ones (lower~$i$) as in a citation network.  (Although we use
the language of time in this paper, the ordering does not have to be a time
ordering.  In a food web, for example, the ordering represents trophic
level.)

It is not possible to construct an acyclic network on every degree
sequence.  Degree sequences, for instance, in which the first vertex has
any outgoing edges ($\kout_1>0$) will not work because there are no earlier
vertices for those edges to attach to.  More generally, all edges outgoing
from vertices $1$ to $i$ must attach at their other end to vertices in the
range $1$ to $i-1$ and hence a necessary condition on the degree sequence
is $\sum_{j=1}^{i-1} \kin_j \ge \sum_{j=1}^i \kout_j$ for all~$i$, with the
inequality becoming an equality for $i=1$ and $i=n$.  Defining the useful
quantity
\begin{equation}
\lambda_i = \sum_{j=1}^{i-1} \kin_j - \sum_{j=1}^i \kout_j,
\end{equation}
this condition can also be written as $\lambda_i\ge0$ for~$i=2\ldots n-1$
and $\lambda_1=\lambda_n=0$.  It is straightforward to prove that this is
also a sufficient condition for a degree sequence to be realizable as a
network.  Physically, $\lambda_i$~represents the number of edges that go
around vertex~$i$, meaning the number that connect vertices later than~$i$
to vertices earlier than~$i$.

We can visualize the degree sequence as a set of edge ``stubs,'' outgoing
and ingoing, attached in the appropriate numbers to each vertex.  Our job
is to match these stubs in pairs to create directed edges.  Our definition
of a random graph for directed acyclic networks is analogous to that of the
standard ``configuration model'' for undirected
networks~\cite{Bollobas80,MR95,NSW01}: it is the graph generated by drawing
uniformly at random from all allowed matchings of the stubs, where
``allowed'' in this case means matchings that respect the ordering of the
vertices.  More correctly it is the \emph{ensemble} of such matchings in
which each matching appears with equal probability.  Note that, as in other
random graph models, multiedges are allowed, although in general they
constitute a fraction only $\Ord(1/n)$ of all edges and hence are usually
negligible.

An attractive feature of this model is that there turns out to be a simple
and efficient algorithm for generating the networks.  Previous numerical
schemes for generating acyclic graphs have relied on Monte Carlo
techniques~\cite{MDB00,IC02}, which are effective but slow.  Our model, by
contrast, allows a simple constructive algorithm: starting with no edges in
our network, we go through each vertex in time order and attach each
outgoing stub to an ingoing stub at an earlier vertex, chosen uniformly at
random from the set of such stubs that are currently unattached.  With a
suitable choice of data structures this algorithm runs in time of order the
number of edges in the network.  In practice, we can easily generate
networks of up to a few billion vertices in reasonable running times.

It may not be immediately clear that this algorithm generates networks with
the same probabilities as the model defined above, but it is easily proved.
Consider the step of the algorithm in which we choose the destinations of
the $\kout_i$ outgoing stubs at vertex~$i$.  At the start of this step, the
number of unused ingoing stubs at earlier vertices is $\sum_{j=1}^{i-1}
\kin_j - \sum_{j=1}^{i-1} \kout_j = \lambda_i + \kout_i$, and the number of
distinct matchings of $i$'s outgoing stubs to these ingoing ones is
$N_i=(\lambda_i+\kout_i)!/\lambda_i!$, each of which has the same
probability $1/N_i$ of being chosen.  Thus the total probability of
generating a specific matching for the whole network is $\prod_{i=2}^n(1/N_i)$, which is clearly uniform over all matchings, as required, since it depends only on the degree sequence and not on the matching itself.

Having defined our model and a method for drawing from its ensemble, we
turn to the calculation of its properties.  Our first goal is to find one
of the most fundamental of network quantities, the probability of
connection between a given pair of vertices, or more correctly the expected
number of edges between them.  Let us define $f_{ij}$ to be the probability
of connection between a given in-stub at vertex~$i$ and a given out-stub at
vertex~$j$, multiplied by the total number~$m$ of edges in the network.
The stub connection probability is equal to the number of complete
matchings in which these particular stubs are connected divided by the
total number of matchings.  Assuming $i<j$, this gives
\begin{equation}
f_{ij} = m\, {\prod_{l=i+1}^{j-1} \lambda_l\over
              \prod_{l=i+1}^{j} (\lambda_l+\kout_l)}.
\label{eq:pij}
\end{equation}
Then the expected number~$P_{ij}$ of edges between $i$ and $j$ is
\begin{equation}
P_{ij} = {\kin_i \kout_j\over m}f_{ij}.
\label{eq:pijsoln}
\end{equation}
Note that in an ordinary (cyclic) directed random graph the expected number of edges between two vertices is $\kin_i\kout_j/m$ and hence $f_{ij}$ is the factor
by which that number is modified in our acyclic model.

By suitable manipulation, Eq.~\eqref{eq:pij} can be rewritten as a product
of independent functions of $i$ and~$j$: $f_{ij}=f_{1n} a_i b_j$, with
$a_1=b_n=1$ and
\begin{equation}
a_i = \prod_{l=2}^i \biggl[ 1 + {\kout_l\over\lambda_l} \biggr],\qquad
b_j = \prod_{l=j}^{n-1} \biggl[ 1 + {\kin_l\over\lambda_l} \biggr]
\label{eq:defsfg}
\end{equation}
for all other $i,j$.  This reduces the calculation of $P_{ij}$ to the
calculation of just $\Ord(n)$ quantities, and for numerical purposes this
is the quickest way to evaluate~$P_{ij}$.  Equation~\eqref{eq:defsfg} also
has the virtue of being manifestly symmetric with respect to in- and
out-degrees (by contrast with Eq.~\eqref{eq:pij}).

\begin{figure}
\begin{center}
\includegraphics[width=8cm]{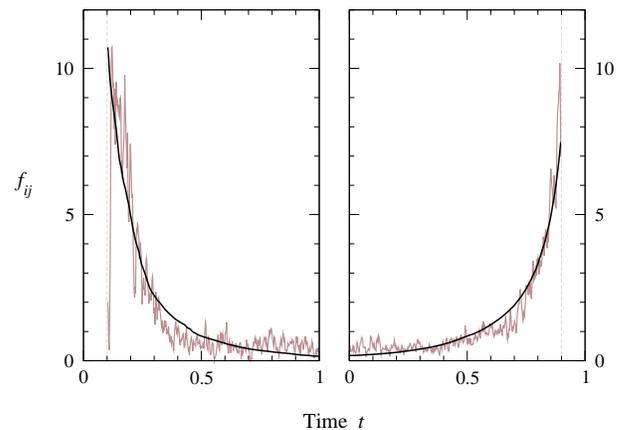}
\end{center}
\caption{Comparison of empirical measurements (jagged lines) and analytic
  predictions (curves) of $f_{ij}$ for the citation network described in
  the text.  The ``time'' of paper~$i$ is defined to be~$t=i/n$.  Left:
  $f_{ij}$~for citations to papers at time $0.1$ (dotted line) from later
  times~$t$.  Right: $f_{ij}$~for citations from papers at time $0.9$ to
  earlier times~$t$.}
\label{fig:citation}
\end{figure}

As a demonstration of the application of the model, we show in
Fig.~\ref{fig:citation} a comparison of our theoretical predictions for
$f_{ij}$ with measured values for a citation network consisting of
$n=27\,221$ physics papers on high-energy theory posted on the Physics
E-print Archive at arxiv.org between 1992 and 2003.  We study $f_{ij}$
rather than $P_{ij}$ since the latter is strongly dependent on the degrees
of individual vertices, via Eq.~\eqref{eq:pijsoln}, making it a noisy
function of its indices.  By contrast, $f_{ij}$~has only a weak dependence
on individual degrees and is relatively smooth.  We estimate $f_{ij}$ for
the observed network by counting the number of edges running between two
windows of width 200 vertices centered on $i$ and~$j$, dividing by the
number of in-stubs in the first window and out-stubs in the second, and
multiplying by~$m$.

As the figure shows, theory and observation are in remarkably good
agreement in this case, indicating that the edge probabilities are, at
least on average, not far from those of the random graph.  A normal (not
acyclic) random directed graph~\cite{NSW01}, sometimes used as a crude
model for acyclic networks, would have $f_{ij}=1$ for all $i,j$---a
perfectly horizontal line in the figure---which would be entirely
incompatible with the observations.  (Other models, particularly
preferential attachment models~\cite{BA99b,Price76}, make quite good models
of citation networks, but our model is more general, being applicable also
to many other acyclic networks for which preferential attachment is not a
good match.)

To make further progress it is convenient to consider, as with other random
graph models, the behavior of the model in the limit of large network size.
Let us define a ``time'' variable~$t\in(0,1]$ such that the time of
vertex~$i$ is $t=i/n$, and let $\kappain(t)$ and $\kappaout(t)$ be the
densities of ingoing and outgoing edges over time, meaning that
$\kappain(t)\>\d t$ is the fraction of ingoing edges in the interval $t$ to
$t+\d t$, and similarly for~$\kappaout(t)$.  By analogy with earlier
developments we also define
\begin{equation}
\lambda(t) = \int_0^t \bigl[ \kappain(t') - \kappaout(t') \bigr] \>\d t',
\end{equation}
and we define $f(t,u)$ to be $m$ times the probability that an in-stub at
time~$t$ is connected to an out-stub at time~$u$.  Then, taking
$n\to\infty$ in Eq.~\eqref{eq:defsfg} and assuming that $\lambda_i$ is large
compared to individual degrees, we find that $f(t,u) = f(0,1)\,a(t)b(u)$,
where
\begin{equation}
a(t) = \exp\biggl[ \int_0^t\! {\kappaout(t')\over\lambda(t')} \d t'
           \biggr],\quad
b(u) = \exp\biggl[ \int_u^1\! {\kappain(u')\over\lambda(u')} \d u' \biggr].
\end{equation}
Since every out-stub must connect to \emph{some} in-stub, $f(t,u)$ must
also satisfy the normalization condition $\int_0^u \kappain(t) f(t,u)
\,\d t = 1$.  Substituting for $f(t,u)$ from above and setting
$u=1$ then gives
\begin{equation}
f(0,1) = \biggl[ \int_0^1 \kappain(t) a(t) \>\d t \biggr]^{-1},
\end{equation}
which allows us to determine the overall normalization of~$f(t,u)$.  If we
wish we can also translate these results back into the language of
individual vertices and write the probability of connection between
vertices $i$ and~$j$ as $P_{ij} = \kin_i\kout_j f(i/n,j/n)/m$.

As an example, consider a random acyclic graph with
\begin{equation}
\kappain(t) = 2(1-t),\qquad
\kappaout(u)=2u.
\label{eq:cascade1}
\end{equation}
Using the formulas above, we then find that
\begin{equation}
f(t,u) = {1\over2(1-t)u}.
\label{eq:ptu}
\end{equation}
Note that this diverges at $t=1$ and $u=0$, as it should: the probability
of connection between an out-stub at time~$u$ and an earlier in-stub
becomes large when $u$ approaches zero because the number of earlier
in-stubs is small (and similarly when $t$ is large).

The probability of connection between \emph{vertices} on the other hand
does not diverge.  Multiplying~\eqref{eq:ptu} by $\kin_i\kout_j/m$ with
$i=nt$, $j=nu$, averaging over the distributions of the degrees, and noting
that the average in- and out-degrees at time $t$ are $c\kappain(t)$ and
$c\kappaout(t)$ where $c=m/n$ is the average degree (in or out) of the
network as a whole, we get
\begin{equation}
P_{ij} = {c\kappain(t)c\kappaout(u)\over m}f(t,u) 
       = {2c(1-t)\times2cu\over2m(1-t)u} = {2c\over n},
\label{eq:cascade}
\end{equation}
which is constant.  Thus all pairs of vertices are equally likely to be
connected.  In fact, this case is closely related to the so-called cascade
model, an acyclic graph model used in the study of food webs~\cite{CN85}.
The cascade model also has constant probabilities of connection between
vertices and moreover it can be shown that all networks with a given degree
sequence appear with the same probability in the cascade model, so that the
set of such networks is a random acyclic graph in our sense~\cite{note1}.

As another example consider the widely studied class of networks generated
by linear preferential attachment processes~\cite{BA99b,KR01,DM02,Price76}.
Because of the inherent time-ordering of their vertices, these processes
generate directed acyclic graphs and are commonly used as a simple model
for citation networks among other things~\cite{Price76}.

For a preferential attachment model in which each vertex created has
out-degree~$c$ and attachment is proportional to $\kin_i+r$ with $c$ and
$r$ constants, the mean in-degree as a function of time is
$r(t^{-c/(c+r)}-1)$~\cite{KR01,DM02}.  Consider a random acyclic graph with
the same in- and out-degrees.  Using the formulas above, we find that
\begin{equation}
f(t,u) = {1\over(1+r/c)(1-t^{c/(c+r)})u^{r/(c+r)}},
\end{equation}
which again diverges at $t=1$ and $u=0$.  The average probability of
connection between vertices $i$ and $j$ is then
\begin{equation}
P_{ij} = {cr\over c+r}\,i^{-c/(c+r)} j^{-r/(c+r)}.
\label{eq:prefattach}
\end{equation}
Remarkably, this is precisely the connection probability for the original
preferential attachment model itself~\cite{DM02}.  Indeed it can be shown,
as with the cascade model, that networks with a given degree sequence occur
with uniform probability in the preferential attachment model, and hence
form a random acyclic graph according to our definition of the term.  It is
sometimes claimed that graphs generated by the preferential attachment
process are not truly random, since they contain correlations of various
kinds~\cite{KR01}.  Our results indicate, however, that, when one correctly
accounts for the time ordering of the vertices, the preferential attachment
model is in fact simply a random graph.

There are many other properties that can be computed for our model.
Consider, for example, the number of paths between vertices in the network.
Let $D_{ij}$ be the expected number of directed paths from $j$ to~$i$.
Since every such path consists either of just a single edge from $j$ to~$i$
or of a path from $j$ to some intermediate vertex~$v$ and then an edge from
$v$ to~$i$, we can write
\begin{equation}
D_{ij} = P_{ij} + \sum_{v=i+1}^{j-1}P_{iv}D_{vj}.
\end{equation}
After some computation, we then find that
\begin{equation}
D_{ij} = P_{ij} \prod_{v=i+1}^{j-1}
         \biggl[ 1+{\kin_v \kout_v\over\lambda_v} \biggr].
\end{equation}

\begin{figure}[t]
\includegraphics[width=\figurewidth]{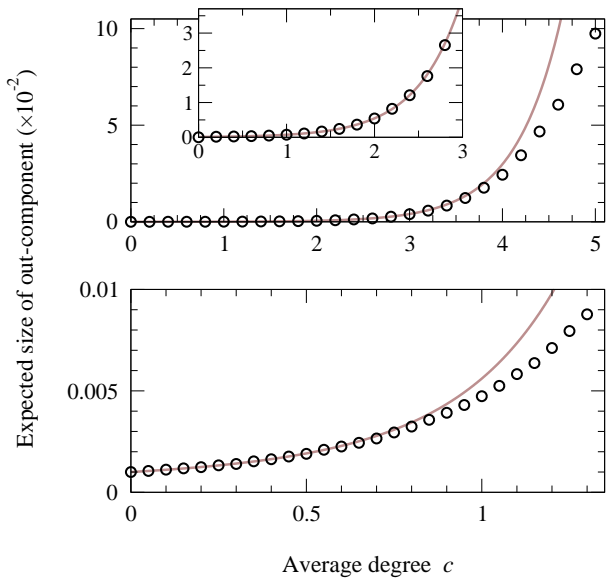}
\caption{Expected size of the out-component for the last ($t=1$) vertex in
  a graph measured as a fraction of system size.  Solid lines represent the
  theoretical predictions.  Points represent numerical results, averaged
  over 8000 graphs.  Top: networks with the degree distribution of the
  cascade model.  Inset: an enlargement of the leftmost portion of the
  curve, showing the agreement between theory and simulation in this
  region.  Bottom: networks with the degree distribution of the
  preferential attachment model with $r=\frac12 c$.}
\label{fig:out}
\end{figure}

When $D_{ij}$ is small, so that the probability of having more than one
path is negligible, $D_{ij}$~can be treated as the probability that a path
exists.  Within this ``tree-like'' regime, we can compute various
quantities of interest starting from the expression for~$D_{ij}$.  For
instance, let $s_j$ be the average size of the out-component reachable from
vertex~$j$---the total number of papers cited directly or indirectly by~$j$
in the language of citation networks.  Then $s_j = 1+\sum_{i=1}^{j-1}
D_{ij}$, which can be evaluated explicitly in the large graph size limit.
For the case of a cascade-type model obeying Eq.~\eqref{eq:cascade1}, for
example, this expression gives $s(t)=\e^{2ct}$, increasing exponentially
with time and largest for the last vertex in the network.  The tree-like
assumption breaks down if $D_{ij}>\Ord(1/n)$ or equivalently if the sizes
of out-components approach the size of the entire network.  For the cascade
model this happens if $\e^{2c}\sim n$, or equivalently $c\sim\frac12\ln n$.
Hence this breakdown is effectively a finite-size effect---in the limit of
large~$n$ it is never observed.  For other choices of degrees, however, the
assumption of tree-like components can break down even in the large~$n$
limit.  The preferential-attachment-type network is an example of this; here the assumption breaks down at $c=1$.  Figure~\ref{fig:out} shows
a comparison of simulations and theory for both cases as a function of~$c$.
Agreement is excellent until we approach the expected breakdown point, at
which simulation and theory diverge significantly.

In conclusion, we have proposed a random graph model for directed acyclic
graphs, a large and important class that describes many real-world
networks.  We have defined the model for arbitrary degree sequences, given
a fast algorithm for generating networks drawn from the model, and shown
that a variety of the model's properties can be calculated exactly, both at
finite sizes and in the limit of large network size.  Just as ordinary
undirected and directed random graphs have played many roles in the
development of network theory, so the acyclic equivalent should prove
useful in the study of acyclic networks, providing an analytically
tractable model for structural network properties, a starting point for
more complex analytic or numerical models, a null model for statistical
comparisons, and, we hope, other applications not yet envisioned.

This work was funded in part by the National Science Foundation under grant
DMS-0804778 and by the James S. McDonnell Foundation.


\begin{thebibliography}{21}
\bibitem{note1}
  There is in fact a slight difference between the cascade model and the
  random acyclic graph in that the former forbids multiedges while the
  latter does not.
\end{thebibliography}
\end{document}